\newcommand{\beq}{\begin{equation}}
\newcommand{\eeq}{\end{equation}}
\newcommand{\bea}{\begin{eqnarray}}
\newcommand{\eea}{\end{eqnarray}}
\newcommand{\half}{{\scriptstyle{{1\over 2}}}}

\def\eg{{\it e.g.}}
\def\ie{{\it i.e.}}
\def\order{{\cal O}}

\def\abinit{{\it ab initio}}
\def\lt{{\mathaccent "7E \lambda}}

\def\h0{{\cal H}^\Lambda_0}

\def\H{{\cal H}^\Lambda}
\def\V{{\cal V}^\Lambda}
\def\Vbar{{\overline V}}

\documentclass{ws-p8-50x6-00}

\begin{document}

\title{Light-Front Quantum Chromodynamics}

\author{Robert J. Perry}

\address{Physics Department, The Ohio State University, Columbus, OH 43210,
USA\\E-mail: perry.6@osu.edu}


\maketitle

\abstracts{ Light-front quantum chromodynamics may lead to an accurate
constituent approximation for the low-energy properties of hadrons. This
requires a cutoff that violates explicit gauge invariance and Lorentz
covariance, leading to the calculation of a renormalized QCD hamiltonian
using a similarity renormalization group. Renormalization repairs broken
symmetries, and in light-front field theory it moves dynamical effects that
usually require large numbers of partons to few-body effective interactions.
This has been shown to work in QED through dominant contributions to the Lamb
shift. In QCD logarithmic confinement arises at second order, and initial
bound state calculations produce reasonable results.}

%%%%%%%%%%%%%%%%%%%%%%%%%%%%%%%%%%%%%%%%%%%%%%%%%%%%%%%%%%%%%%%%%%%%%%%%%%

\section{Motivation and Outline}

Quantum chromodynamics (QCD) is the fundamental theory of the strong
interaction, but our understanding of QCD and our ability to use it to solve
low-energy problems where the interactions are truly strong falls far short
of our accomplishments in the study of quantum electrodynamics (QED). Both
are gauge theories, but the fact that gluons carry color charge drastically
complicates QCD. Couplings in QED are weak at low energies and photons are
nearly free, so that the interactions their exchange produces are readily
approximated. As a result low energy bound states can be accurately
described using a small number of constituents whose interactions appear in
the hamiltonian at second order, and the vacuum has no effect on bound
states in light-front QED. 

The study of QCD and strong interaction phenomenology leads to little hope of
deriving similar approximations. Interactions in QCD are strong at
low energies and the color-charged gluons interact strongly while mediating
interactions. The vacuum is supposed to have a complicated structure to
which conferences are devoted. Its structure is assumed to be responsible
for essential aspects of the theory such as confinement and chiral symmetry
breaking in all widely respected treatments of the theory, and it is hard to
imagine an accurate description of individual hadrons that does not also
include a complicated description of the vacuum in which they reside.
Nonetheless, we advocate an approach in which an accurate
description of hadrons resembles the accurate description of atoms in
QED.\cite{nonpert,brazil}

We start with the heretical conjecture that {\it a constituent picture
of hadrons can be derived from QCD}. This conjecture guides our calculations,
but the approach I describe is completely fixed by QCD and the
constituent picture will fail if it is inadequate.\footnote{The
description of light hadrons requires refinements to the simple
approximations I describe, because of chiral symmetry breaking.}

{\it If} a constituent approximation is accurate, we can study the low-energy
properties of hadrons ({\it e.g.}, mesons) by solving a relativistic
Schr{\"o}dinger equation:

\begin{equation}
H_\Lambda \mid \Psi_\Lambda\rangle = E \mid \Psi_\Lambda \rangle,
\end{equation}
with,
\begin{equation}
\mid \Psi_\Lambda \rangle = \phi^\Lambda_{q\bar{q}} \mid q\bar{q}
\rangle + \phi^\Lambda_{q\bar{q}g} \mid q\bar{q}g \rangle +
\cdot\cdot\cdot \;,
\end{equation}
where I use shorthand notation for the Fock space components of the state. 
The exact state vector includes an infinite number of terms. In a constituent
approximation we truncate this series, adding terms to improve the
approximation.  We derive the hamiltonian from QCD, so we must allow for the
possibility of constituent gluons.  I have indicated that the hamiltonian and
the state both depend on a cutoff, $\Lambda$, which is critical for the
approximation.

This approach has no chance of working without a {\it renormalization scheme
tailored to light-front hamiltonian field theory.}  Much of our work has
focused on the development of such a renormalization
scheme.\cite{wilgla}$^-$\cite{brent}  

Consider the conditions under which it
might be possible to truncate the above Fock space series without making an
arbitrarily large error in the eigenvalue.  I focus on the eigenvalue,
because it is certainly not possible to approximate all observable
properties of hadrons (\eg, wee parton structure functions) this way.
For this approximation to be valid, {\it all} many-body states must
approximately decouple from the dominant few-body components.

We know that
even in perturbation theory, high energy many-body states do not decouple
from few-body states.  In fact, the errors from simply discarding high
energy states are infinite.  In second-order perturbation theory, for
example, high energy photons contribute an arbitrarily large shift to the
mass of an electron.  This second-order effect is illustrated in Fig. 1. 
The solution to this problem is well-known, renormalization. 
Renormalization moves the effects of high energy components in the state to
effective interactions in the hamiltonian.

\begin{figure}
\epsfxsize=20pc
\center{\epsfbox{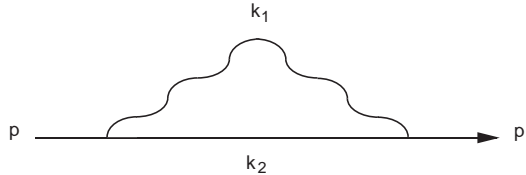}}
\caption{The second-order shift in the self-energy of a bare electron
due to mixing with electron-photon states.}
\label{fig: fig1}
\end{figure}

It is difficult to see how a constituent approximation can emerge in a
hamiltonian calculation using any regularization scheme without a cutoff
that either removes high energy degrees of freedom or removes direct
couplings between low and high energy degrees of freedom.

In the best case scenario we expect the cutoff to act like a resolution.  If
the cutoff is increased to an arbitrarily large value, the resolution
increases and instead of seeing a few constituents we resolve the
substructure of the constituents and the few-body approximation breaks down.
As the cutoff is lowered, this substructure is removed from the state
vectors, and the renormalization procedure replaces it with effective
interactions in the hamiltonian.  Any ``cutoff" that does not remove this
substructure from the states is of no use to us.  

This point is well-illustrated by the QED calculations discussed
below.\cite{brazil,BJ97a,BJ97c} There is a window into which the cutoff must
be lowered for the constituent approximation to work.  If the cutoff is
raised atomic states start to include photons, and as the cutoff is raised
further they start to include additional photons and electron-positron
pairs.  After the cutoff is lowered to a value that can be self-consistently
determined {\it a-posteriori}, photons and pairs are removed from the states
and replaced by the Coulomb interaction and relativistic corrections in the
hamiltonian.  The cutoff cannot be lowered too far using a perturbative
renormalization group, hence the window.

Thus, if we remove high energy degrees of freedom, or coupling to high energy
degrees of freedom, we should encounter self-energy corrections leading to
effective one-body operators, vertex corrections leading to effective
vertices, and  exchange effects leading to explicit many-body interactions
not found in the canonical hamiltonian.  We naively expect these operators
to be local when acting on low energy states,  because simple uncertainty
principle arguments indicate that high energy virtual particles cannot
propagate very far.  Unfortunately these arguments break down in light-front
coordinates, and at best we can maintain transverse locality.\cite{perryrg}

\begin{figure}
\epsfxsize=20pc
\epsfbox{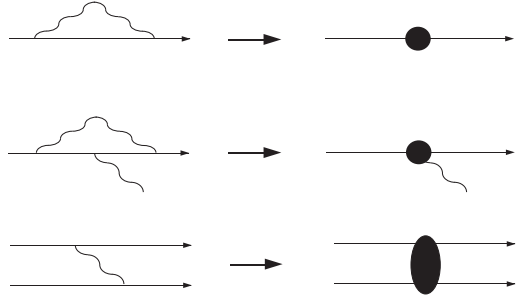}
\caption{Mixing of low-energy few-body states with high-energy many-body
states alters the dispersion relation for single particles, alters emission
and absorption vertices, and produces approximately local few-body
interactions.}
\label{fig:fig2}
\end{figure}

In this article I give a brief overview of the new renormalization techniques
we employ, and in a separate article in these Proceedings \cite{japan_rg}
S{\'e}rgio Szpigel and I illustrate them using the two-dimensional
$\delta$-function potential.

Low energy many-body states do not typically decouple from low energy
few-body states.  The worst of these low energy many-body states is the
vacuum.  This is what drives us to use light-front coordinates.\cite{dirac}
Fig. 3 shows a pair of particles being produced out of the vacuum in
equal-time coordinates $t$ and $z$.  The transverse components $x$ and $y$
are not shown, because they are the same in equal-time and light-front
coordinates.  The figure also shows the light-front time axis,
\begin{equation}
x^+=t+z\;, 
\end{equation}
and the light-front longitudinal spatial axis,
\begin{equation}
x^-=t-z\;.
\end{equation}

\begin{figure}
\epsfxsize=16pc
\center{\epsfbox{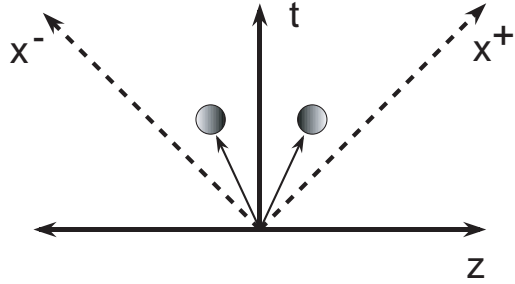}}
\caption{Light-front coordinates. Light-front `time' is $x^+=t+z$, and
light-front longitudinal position is $x^-=t-z$.}
\label{fig:fig3}
\end{figure}

In equal-time coordinates it is kinematically possible for virtual pairs to
be produced from the vacuum, as long as their momenta sum to zero so that
three-momentum is conserved.  Because of this, the state vector for a proton
includes an arbitrarily large number of particles that are disconnected from
the proton.  The only constraint imposed by relativity is that particle
velocities be less than or equal to that of light.

In light-front coordinates, however, we see that all allowed
trajectories lie in the first quadrant.  In other words,
light-front longitudinal momentum, $p^+$ (conjugate to $x^-$
since
$a\cdot b=\half (a^+ b^- + a^- b^+) - {\bf a}_\perp \cdot {\bf
b}_\perp$), is always positive,
\begin{equation}
p^+ \ge 0 \;.
\end{equation}
We exclude particle modes with $p^+=0$, forcing the vacuum to be trivial
because it is the only state with $p^+=0$.  Moreover, the light-front energy
of a free particle of mass $m$ is
\begin{equation}
p^-={{\bf p}_\perp^2+m^2 \over p^+} \;.
\end{equation}
This implies that all free particles with zero longitudinal momentum
have infinite energy, unless their mass and transverse momentum are zero.

Is the vacuum really trivial? What about confinement? What about chiral
symmetry breaking? What about instantons? What about the job security of
theorists who study the vacuum?

I simply discard all $p^+=0$ degrees of freedom and replace their
effects using a renormalization procedure that does not require any explicit
reference to zero modes. Thus the vacuum in our formalism is trivial.  We are
forced to work in the ``hidden symmetry phase" of the theory, and to
introduce effective interactions that reproduce all effects associated with
the vacuum in other formalisms.\cite{nonpert,burkardt93,burkardt97} The
simplest example of this approach is provided by a scalar field theory with
spontaneous symmetry breaking.  It is possible to shift the scalar field and
deal explicitly with a theory containing symmetry breaking interactions.  In
the simplest case $\phi^3$ is the only relevant or marginal symmetry breaking
interaction, and one can simply tune this coupling to the value
corresponding to spontaneous rather than explicit symmetry breaking.

The use of a symmetry-breaking cutoff and the removal of zero-modes leads to
a large number of operators not found in the canonical QCD hamiltonian. This
complicates the renormalization procedure, but it may lead to tremendous
simplifications in the final nonperturbative problem.  For example, few-body
operators must produce confinement manifestly! 

Confinement cannot require particle creation and annihilation, flux tubes,
etc.  This is easily seen using a variational argument.  Consider a color
neutral quark-antiquark pair that are separated by a distance $R$, which is
slowly increased to infinity.  Moreover, to see the simplest form of
confinement assume that there are no light quarks, so that the energy should
increase indefinitely as they are separated if the theory possesses
confinement.  At each separation the gluon components of the state adjust
themselves to minimize the energy.  But this means that the expectation
value of the hamiltonian for a state with no gluons must exceed the energy
of the state with gluons, and therefore must diverge even more rapidly than
the energy of the true ground state.  This means that there must be a
two-body confining interaction in the hamiltonian.  If the renormalization
procedure is unable to produce such confining two-body interactions, it
is invalid.

\subsection{Simple Strategy}

I want to outline a conceptually simple strategy for bound state
calculations.\cite{brazil}  The first step is to use a perturbative
similarity renormalization group \cite{wilgla,wegner} and coupling
coherence \cite{coupcoh,perryrg} to find the renormalized hamiltonian as an
expansion in powers of the canonical coupling:
\begin{equation}
H^\Lambda = h_0 + g_\Lambda h_1^\Lambda + g_\Lambda^2 h_2^\Lambda +
\cdot \cdot \cdot \;.
\end{equation}
We compute this series to a finite order and to date have not required any
{\it ad hoc} assumptions to uniquely fix the hamiltonian.  No operators are
added to the hamiltonian by hand, so it is completely determined
by the underlying theory to this order. This step is illustrated in a
separate article in these Proceedings.\cite{japan_rg}

The second step is to employ bound state perturbation theory to solve the
eigenvalue problem.  The complete hamiltonian contains every interaction
(although each is cut off) contained in the canonical hamiltonian, and many
more.  We separate the hamiltonian,
\begin{equation}
H^\Lambda=\h0+\V \;,
\end{equation}
treating $\h0$ nonperturbatively and computing the effects of
$\V$ in bound state perturbation theory. We must choose
$\h0$ and $\Lambda$ so that $\h0$ is tractable and to
minimize corrections from higher orders of $\V$ within a
constituent approximation.

If a constituent approximation is valid {\it after} $\Lambda$ is lowered to a
critical value that must be determined, we may be able to move all particle
creation and annihilation to $\V$.  $\h0$ includes
many-body interactions that do not change particle number, and these
interactions should be primarily responsible for the constituent bound state
structure.

There are several obvious flaws in this strategy. Chiral symmetry-breaking
operators, which must be included in the hamiltonian since we work entirely
in the hidden symmetry phase of the theory, do not appear at any finite
order in the coupling.  There is only one relevant chiral symmetry breaking
operator, and it appears in the canonical hamiltonian when quarks
are massive (spin-flip gluon emission by quarks) although it can acquire
non-canonical dependence on longitudinal momenta since there is no
longitudinal locality. This operator must simply be added if quarks are
massless and tuned to fit spectra or fixed by a non-perturbative
renormalization procedure.\cite{nonpert,mustaki,Br94b} In addition, there
are perturbative errors in the strengths of all operators. 
We know from simple scaling arguments \cite{wilson75} that when $\Lambda$ is
in the perturbative scaling regime:
\begin{itemize}
\item{small errors in relevant operators exponentiate in the output,}
\item{small errors in marginal operators produce comparable errors in
output,}
\item{small errors in irrelevant operators tend to decrease
exponentially in the output.}
\end{itemize}
This means that even if a relevant operator appears (\eg, a constituent quark
or gluon mass operator), we may need to tune its strength rather than use
its perturbative value to obtain reasonable results.  We have not had to do
this, but we have recently studied some of the effects of tuning a gluon mass
operator.\cite{Sz97a}

To date this strategy has produced well-known results in QED
\cite{brazil,BJ97a,BJ97c} through the Lamb shift, and reasonable results for
heavy quark bound states in QCD.\cite{Br96a,Br97a,Sz97a} All of these
calculations rely on a nonrelativistic reduction of the effective
hamiltonian, which leads to drastic simplifications. Glueball calculations
are being completed \cite{brent2} and will be the first fully
relativistic QCD calculations in this approach.

The best place to begin the study of light-front field theory and to see how
a constituent approximation can arise in light-front gauge theories is the
Schwinger model, massless QED in $1+1$ dimensions.\cite{schwing} This model
can be solved analytically,\cite{lowenstein} and its physical content is
remarkably simple. There is only one physical particle, a massive neutral
scalar particle with no self-interactions. The Fock space content of the
physical states depends crucially on the coordinate system and gauge, and it
is only in light-front coordinates that a simple constituent picture
emerges.\cite{bergknoff,brazil} In light-front field theory the
physical particle is a bound state of a single electron-positron pair, with
a wave function that is constant in the longitudinal fraction carried by
either particle.

The Schwinger model does not display the renormalization problems that must
be solved in QED$_{3+1}$ and QCD$_{3+1}$. Before turning to these theories,
I discuss the renormalization machinery that has been developed for
light-front hamiltonians.

%%%%%%%%%%%%%%%%%%%%%%%%%%%%%%%%%%%%%%%%%%%%%%%%%%%%%%%%%%%%%%%%%%%%%%%%%%

\section{Light-Front Renormalization Group}

In $3+1$ dimensions we must introduce a cutoff, $\Lambda$,
and we never perform explicit bound state calculations with $\Lambda$
anywhere near its continuum limit. In fact, we want to let $\Lambda$ become
as small as possible.  In my opinion, any strategy for solving light-front
QCD that requires the cutoff to explicitly approach infinity in the
nonperturbative part of the calculation is useless.  Therefore, we must set
up and solve
\begin{equation}
P^-_\Lambda \mid \Psi_\Lambda(P) \rangle = {{\bf P}_\perp^2 + M^2
\over P^+} \mid \Psi_\Lambda(P) \rangle \;.
\end{equation}

Physical results, such as the mass, $M$, can not depend on the
arbitrary cutoff, $\Lambda$.  This means that $P^-_\Lambda$ and $\mid
\Psi_\Lambda\rangle$ must depend on the cutoff in such a way that $\langle
\Psi_\Lambda \mid P^-_\Lambda \mid \Psi_\Lambda \rangle$ does not.  Wilson
based the derivation of his renormalization group on this
observation,\cite{wilson65,wilson70,wilson74,wilson75} and we modify Wilson's
renormalization group to compute $P^-_\Lambda$.

It is difficult to even talk about how the hamiltonian depends on the
cutoff without having a means of changing the cutoff.  If we can
change the cutoff, we can explicitly watch the hamiltonian's cutoff
dependence change and fix its cutoff dependence by insisting that
this change satisfy certain requirements (\eg, that the limit in
which the cutoff is taken to infinity exists).  We introduce an
operator that changes the cutoff,
\begin{equation}
H(\Lambda_1) = T[H(\Lambda_0)] \;,
\end{equation}
where I assume that $\Lambda_1 < \Lambda_0$.  To simplify the
notation, I let $H(\Lambda_l)=H_l$. To renormalize the
hamiltonian we study the properties of the transformation.

\begin{figure}
\epsfxsize=20pc
\center{\epsfbox{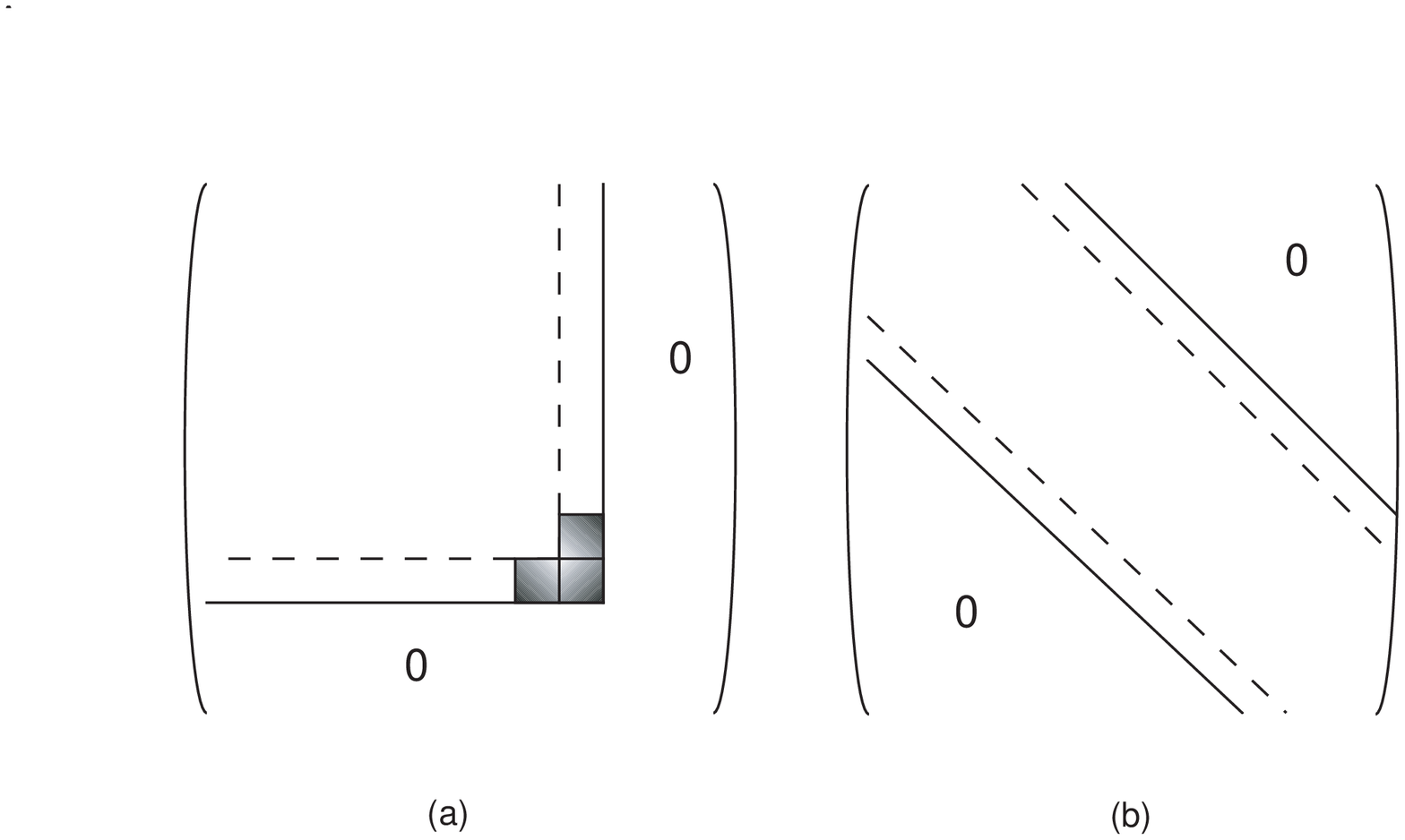}}
\caption{Two ways to run a cutoff on free energy.  In (a) a cutoff
on the magnitude of the energy is lowered from the solid to the
dashed lines, with problems resulting from the removed shaded
region. In (b) a cutoff on how far off diagonal matrix elements
appear is lowered from the dashed to the solid lines.}
\label{fig:fig4}
\end{figure}

Fig. 4 displays two generic cutoffs that might be used.
Traditionally theorists have used cutoffs that remove high energy
states, as shown in Figure 4a.  This is the type of cutoff Wilson
employed in his initial work \cite{wilson65} and I have studied its use
in light-front field theory.\cite{perryrg} When a cutoff on energies
is reduced, all effects of couplings eliminated must be moved to
effective operators.  When these
effective operators are computed perturbatively they involve products
of matrix elements divided by energy denominators.  Expressions
closely resemble those encountered in standard perturbation theory,
with the second-order operator involving terms of the form
\begin{equation}
\delta V_{ij} \sim {\langle \phi_i \mid V \mid \phi_k \rangle \langle
\phi_k \mid V \mid \phi_j \rangle \over \epsilon_i-\epsilon_k}\;.
\end{equation}
This new effective interaction replaces missing couplings, so the
states $\phi_i$ and $\phi_j$ are retained and the state $\phi_k$ is
one of the states removed.  The problem comes from the shaded, lower
right-hand corner of the matrix, where the energy denominator
vanishes for states at the corner of the remaining matrix.  In this
corner we should use nearly degenerate perturbation theory rather than
perturbation theory, but to do this requires solving high energy
many-body problems nonperturbatively before solving the low energy
few-body problems.

An alternative cutoff, which does not actually remove any states and that
can be run by a similarity transformation\footnote{In deference to the
original work I call this a similarity transformation even though in
all cases of interest to us it is a unitary transformation.} is shown in
Fig. 4b.  This cutoff removes couplings between states whose free energy
differs by more than the cutoff.  The advantage of this cutoff is that the
effective operators resulting from it contain energy denominators that are
never smaller than the cutoff, so that a perturbative approximation for the
effective hamiltonian may work well.

Before discussing the similarity transformation, which runs a cutoff on
off-diagonal matrix elements, I need to introduce the classification of
operators as relevant, marginal, and irrelevant. This classification is
not necessary to use the similarity renormalization group, but it is
necessary for the definition of coupling coherence, and it is essential for
a full understanding of the renormalization group. It
starts with the definition of a fixed point, which is a hamiltonian that is
left invariant by the transformation,
\begin{equation}
H^* = T[H^*] \;.
\end{equation}
Obviously a fixed point is a renormalized hamiltonian, because the cutoff
fixed point hamiltonian is identical to the fixed point hamiltonian with
infinite cutoff.

Consider the immediate `neighborhood' of the fixed point, and assume
that the trajectory remains in this neighborhood.  This assumption
must be justified {\it a posteriori}, but if it is true we should write
\begin{equation}
H_l=H^*+\delta H_l \;,
\end{equation}
and consider the trajectory of small deviations $\delta H_l$.

As long as $\delta H_l$ is `sufficiently small,' we can use a
perturbative expansion in powers of $\delta H_l$, which leads us to
consider
\begin{equation}
\delta H_{l+1}= L \cdot \delta H_l + N[\delta H_l] \;.
\end{equation}
Here $L$ is the linear approximation of the full
transformation in the neighborhood of the fixed point, and $N[\delta
H_l]$ contains all contributions to $\delta H_{l+1}$ of $\order(\delta
H_l^2)$ and higher.

The object of the renormalization group calculation is to compute
trajectories and this requires a representation for $\delta H_l$.  The
problem of computing trajectories is one of the most common in
physics, and a convenient basis for the representation of $\delta H_l$
is provided by the eigenoperators of $L$, since $L$ dominates the
transformation near the fixed point.  These eigenoperators and their
eigenvalues are found by solving
\begin{equation}
L \cdot O_m=\lambda_m O_m \;.
\end{equation}
For light-front field theory this
linear transformation is a scaling of the transverse coordinate, the
eigenoperators are products of field operators and transverse derivatives,
and the eigenvalues are determined by the transverse dimension of the
operator. All operators can include both powers and inverse powers of
longitudinal derivatives because there is no longitudinal
locality.\cite{perryrg}

Using the eigenoperators of $L$ as a basis we can represent $\delta
H_l$,
\begin{equation}
\delta H_l = \sum_{m\in R} \mu_{m_l}O_m +\sum_{m\in M} g_{m_l}O_m+
\sum_{m\in I} w_{m_l}O_m \;.
\end{equation}
Here the operators $O_m$ with $m\in R$ are {\it relevant}
(\ie, $\lambda_m>1$), the operators $O_m$ with $m\in M$ are {\it
marginal} (\ie, $\lambda_m=1$), and the operators with $m\in I$ are
{\it irrelevant} (\ie, $\lambda_m<1$).  The motivation behind
this nomenclature is made clear by considering repeated application of
$L$, which causes the relevant operators to grow exponentially, the
marginal operators to remain unchanged in strength, and the irrelevant
operators to decrease in magnitude exponentially.

\subsection{Similarity transformation}

Stan G{\l}azek and Ken Wilson studied the problem of small energy
denominators that typically appear in effective interactions,\cite{wilson70}
and realized that a similarity transformation that runs a different form of
cutoff (as discussed above) avoids this problem.\cite{wilgla}
Independently, Wegner\cite{wegner} developed a similarity transformation
that is easier to use than those first studied by G{\l}azek and Wilson.
Recently Walhout has developed a transformation that may be well-suited to
analytical and higher-order calculations.\cite{walhout}

Many details and a simple example of the use of Wegner's similarity
transformation can be found elsewhere in these Proceedings.\cite{japan_rg}
Here I summarize the details needed to complete QED and QCD calculations
through second order. Consider a hamiltonian, $H_s = h + V_s$, where $h$ is
a free hamiltonian that may contain mass terms. The dependence of $H$ on the
cutoff, where $s$ is the inverse of the cutoff squared, is given by
\begin{equation}
\frac{d H_s}{ds}=[H_s,[H_s,h]
\;.
\end{equation}
This equation is most easily analyzed in terms of matrix elements between
eigenstates of $h$,
\begin{equation}
h |\phi_i\rangle = \epsilon_i |\phi_i\rangle \;.
\end{equation}

A reduced interaction, ${\overline V}_s$, can be defined,
\begin{equation}
V_{sij}= e^{-s\Delta_{ij}^2} \; {\overline V}_{sij}
\label{reduced}
\;,
\end{equation}
\noindent where $\Delta_{ij}=\epsilon_i-\epsilon_j$. A gaussian cutoff
factor that clearly forces the interactions towards the diagonal is
isolated, and the equation for $\Vbar_s$ starts at second order,
\begin{equation}
\frac{d\Vbar_{sij}}{ds}=\sum_k\left(\Delta_{ik}+\Delta_{jk}\right)
\; \Vbar_{sik} \; \Vbar_{skj} \; e^{-2s \Delta_{ik}\Delta_{jk}}
\;,\label{wegnerbigone}
\end{equation}
\noindent
where I use 
$\Delta_{ij}^2-\Delta_{ik}^2-\Delta_{jk}^2=-2\Delta_{ik}\Delta_{jk}$.

This is a first-order differential equation and it is exact. Its solution is
completely determined when a complete set of boundary conditions are
specified. In its simplest form the boundary conditions would be given by
specifying $H_{s_0}$ using a single value of $s_0$ for all matrix elements,
with $s_0=0$ corresponding to an infinite cutoff. As discussed in the simple
$\delta$-function example,\cite{japan_rg} divergences prevent us from using
$s_0=0$ to specify boundary conditions for relevant and marginal operators,
but we are free to choose different values of $s_0$ for different operators
and this freedom is crucial. The boundary conditions are given by coupling
coherence, which I discuss in the next section.

For QED and QCD we can approximate $\Vbar_s$ using a perturbative
expansion in powers of a single running coupling constant,
\begin{equation}
\Vbar_s = g_s \Vbar^{^{(1)}} + g_s^2 \Vbar^{^{(2)}}_s + \cdot \cdot \cdot \;.
\end{equation}
In both QED and QCD the coupling does not begin to run until third order,
and the first order term, $\Vbar^{^{(1)}}$, is given by the canonical
hamiltonian and has no dependence on $s$ other than that of the running
coupling. To second order we have
\begin{equation}
\frac{d \Vbar^{^{(2)}}_{sij}}{ds}=\sum_k\left(\Delta_{ik}+\Delta_{jk}\right)
\; \Vbar^{^{(1)}}_{ik}\; \Vbar^{^{(1)}}_{kj} \; e^{-2s
\Delta_{ik}\Delta_{jk}}
\;.
\end{equation}
Integrating from $s_0$ to $s$, we obtain
\begin{eqnarray}
\Vbar_{sij}^{^{(2)}}&=&\frac{1}{2}\sum_k
\Vbar^{^{(1)}}_{ik}\; \Vbar^{^{(1)}}_{kj} \; 
\left(\frac{1}{\Delta_{ik}}+\frac{1}{\Delta_{jk}}\right)\times\nonumber\\
&&~~~~~~~~~~\times \left[e^{-2 s_0 
\Delta_{ik}\Delta_{jk}}-e^{-2s\Delta_{ik}\Delta_{jk}}\right]
\;.
\label{ham2}
\end{eqnarray}
This simple equation is sufficient to compute the renormalized QED and QCD
hamiltonians through second order the canonical couplings.

In order to use
the concept of fixed points in the similarity renormalization group, we need
to modify the transformation slightly so that it is possible to get a fixed
point with interactions. The free hamiltonian $h$ is obviously a fixed
point, but Eq. (\ref{wegnerbigone}) clearly shows that the linearized
transformation is given by Eq. (\ref{reduced}). If we want to have fixed
points for which
$V_s$ is not zero, we need to modify the transformation so that the gaussian
cutoff factor in Eq. (\ref{reduced}) does not change as $s$ changes. This is
most easily accomplished in the traditional manner by rescaling variables so
that the eigenvalues in Eq. (\ref{reduced}) absorb the change in $s$. For
example, in massless light-front field theory,
\begin{equation}
\epsilon_i = p^-_i = {p_{\perp i}^2 \over p^+_i} \;.
\end{equation}
We can absorb a change in $s$ by rescaling all transverse momenta, and this
leads to an operator classification based on the transverse engineering
dimension of operators. There are other interesting
possibilities,\cite{perryrg} but the result is that transverse local
operators ({\i.e.}, no inverse powers of transverse momenta) that vanish
when all transverse momenta are taken to zero are irrelevant. The
electron-photon and quark-gluon couplings are marginal, except for a
spin-flip piece that is relevant and breaks light-front chiral
symmetry.\cite{nonpert,mustaki}

The interesting new feature of light-front field theory that is not
encountered in equal-time or euclidean field theory is that there is no
longitudinal locality, so that the longitudinal dimension of operators does
not affect their classification. A relevant or marginal operator can contain
a function of longitudinal momentum fractions, and these functions
inevitably appear in renormalized light-front hamiltonians.\cite{perryrg}
This means that there are effectively an infinite number of relevant and
marginal operators in renormalized light-front hamiltonians, and such a
situation is usually regarded as a disaster for renormalization group
treatments because it indicates that there may be an infinite number of free
physical parameters. Ken Wilson and I developed coupling coherence to deal
with this problem.\cite{coupcoh,perryrg}

\subsection{Coupling coherence}

The basic mathematical idea behind coupling coherence was
first formulated by Oehme, Sibold, and Zimmerman.\cite{coupcoh2}
They were interested in field theories where many couplings appear,
such as the standard model, and wanted to find some means of reducing
the number of couplings.

The puzzle that led us to the same results is how to reconcile our knowledge
from covariant formulations of QCD that only one running coupling constant
characterizes the renormalized theory with the appearance of new
counterterms and functions required by the light-front formulation.
What happens in perturbation theory when there are effectively an
infinite number of relevant and marginal operators?  In particular,
does the solution of the perturbative renormalization group equations
require an infinite number of independent counterterms (\ie,
independent functions of the cutoff)? Coupling coherence provides the
conditions under which a finite number of running variables
determines the renormalization group trajectory of the renormalized
hamiltonian. To leading nontrivial orders these conditions are
satisfied by the counterterms introduced to restore Lorentz
covariance in scalar field theory and gauge invariance in light-front
gauge theories. In fact, the conditions can apparently be used to determine
all counterterms in the hamiltonian, including relevant and
marginal operators that contain functions of longitudinal momentum fractions;
and with no direct reference to Lorentz covariance, this symmetry seems to be
restored to observables.\cite{perryrg}

A coupling-coherent hamiltonian is analogous to a fixed point
hamiltonian, but instead of reproducing itself exactly it reproduces
itself in form with a limited number of independent running couplings. 
If $g_\Lambda$ is the only independent coupling in a theory, in a
coupling-coherent hamiltonian {\it all other couplings are invariant
functions of $g_\Lambda$, $f_i(g_\Lambda)$}.  The extra couplings
$f_i(g_\Lambda)$ depend on the cutoff only through their dependence
on the running coupling $g_\Lambda$, and in general we demand
$f_i(0)=0$.  This boundary condition on the dependent couplings is
motivated in our calculations by the fact that it is the combination
of the cutoff and the interactions that force us to add the
counterterms we seek, so the counterterms should vanish when the
interactions are turned off.

Let me illustrate the idea with a simple example with a finite number of
relevant and marginal operators \abinit, and use coupling coherence to
discover when only one or two of these may independently run with the
cutoff.  Such conditions are met when an underlying
symmetry exists, although this is not necessary.

Consider a theory in which two scalar fields interact,
\begin{equation}
V(\phi)={\lambda_1 \over 4!}\phi_1^4+{\lambda_2 \over 4!}\phi_2^4+
{\lambda_3 \over 4!}\phi_1^2 \phi_2^2 \;.
\end{equation}
Under what conditions are there fewer than three
independent running coupling constants?  We can use a simple cutoff on
Euclidean momenta,
$q^2<\Lambda^2$.  Letting $t=\ln(\Lambda/\Lambda_0)$, the
Gell-Mann--Low equations are
\begin{equation}
{\partial \lambda_1 \over \partial t} = 3 \zeta \lambda_1^2 + {1
\over 12} \zeta \lambda_3^2 + \order(2\;{\rm loop}) \;,
\label{lambda1}
\end{equation}
\begin{equation}
{\partial \lambda_2 \over \partial t} = 3 \zeta \lambda_2^2 + {1
\over 12} \zeta \lambda_3^2 + \order(2\;{\rm loop}) \;,
\label{lambda2}
\end{equation}
\begin{equation}
{\partial \lambda_3 \over \partial t} = {2 \over 3}
\zeta \lambda_3^2 +  \zeta \lambda_1 \lambda_3+\zeta \lambda_2
\lambda_3 + \order(2\;{\rm loop}) \;;
\label{lambda3}
\end{equation}
where $\zeta=\hbar/(16\pi^2)$.  It is not important at this
point to understand how these equations are derived.

Assume that there is only one independent variable, $\lt=\lambda_1$,
so that $\lambda_2$ and $\lambda_3$ are functions of $\lt$.  In this
case Eqs. (\ref{lambda2}) and (\ref{lambda3}) become,
\begin{equation}
\Bigl(3 \lt^2+{1 \over 12} \lambda_3^2 \Bigr) \; {\partial \lambda_2
\over \partial \lt} = 3 \lambda_2^2+{1 \over 12} \lambda_3^2 \;,
\end{equation}
\begin{equation}
\Bigl(3 \lt^2+{1 \over 12} \lambda_3^2 \Bigr) \; {\partial \lambda_3
\over \partial \lt} = {2 \over 3} \lambda_3^2 + \lt \lambda_3+
\lambda_2 \lambda_3 \;.
\end{equation}
The only non-trivial solutions are $\lambda_2=\lt$, and either $\lambda_3
= 2 \lt$ or $\lambda_3=6 \lt$.  If $\lambda_3=2\lt$,
\begin{equation}
V(\phi)={\lt \over 4!}\;\bigl(\phi_1^2+\phi_2^2 \bigr)^2 \;,
\end{equation}
and we find the $O(2)$ symmetric theory.  If
$\lambda_3=6\lt$,
\begin{equation}
V(\phi)={\lt \over 2 \cdot 4!}\;\Bigl[\bigl(\phi_1+\phi_2\bigr)^4+
\bigl(\phi_1-\phi_2\bigr)^4\Bigr] \;,
\end{equation}
and we find two decoupled scalar fields.  Therefore,
$\lambda_2$ and $\lambda_3$ do not run independently with the cutoff
if there is a symmetry that relates their strength to $\lambda_1$.

The condition that a limited number of variables run with the cutoff
does not only reveal symmetries broken by the regulator, it may also
be used to uncover symmetries that are broken by the vacuum.  For example, it
is straightforward to show that in a scalar theory with a $\phi^3$ coupling,
this coupling can be fixed as a function of the $\phi^2$ and $\phi^4$
couplings only if the symmetry is spontaneously broken rather than explicitly
broken.\cite{coupcoh}

For the QED and QCD calculations, I need to compute
the hamiltonian to second order, while the canonical coupling runs at
third order.  In this case we can use Eq.(\ref{ham2}), with $s_0 \rightarrow
\infty$ for relevant operators and $s_0 \rightarrow 0$ for irrelevant
operators, and with the bare coupling
$e$ replaced by the running coupling $e_s$ or $e_\Lambda$. The more
interesting case of marginal operators can be avoided at
first.\cite{perryrg,brent}

%%%%%%%%%%%%%%%%%%%%%%%%%%%%%%%%%%%%%%%%%%%%%%%%%%%%%%%%%%%%%%%%%%%%%%%%%%

\section{Light-Front QED and QCD}

Various forms of the canonical light-front QED and QCD hamiltonians can be
found in several articles.\cite{brodsky,Zh93a} Following
Brodsky and Lepage, I have displayed these hamiltonians
elsewhere.\cite{brazil,cambridge}

In light-cone gauge and using light-front coordinates, it is possible to
explicitly eliminate all unphysical degrees of freedom and write the
hamiltonian in terms of two-component fermions and transverse
gluons. Any ambiguities in the procedure that come from the zero-mode
problem or normal-ordering are resolved by coupling coherence, so the
renormalized hamiltonian is apparently uniquely determined order-by-order in
the running coupling.

\subsection{Light-front QED}

In this section I follow the strategy outlined in the first
section to compute the positronium spectrum.  I outline the
calculation through the leading order Bohr results \cite{brazil} and
indicate how higher order calculations proceed.\cite{BJ97a,BJ97c} 

The first step is to compute a renormalized cutoff hamiltonian as a
power series in the running coupling $e_\Lambda$,
\begin{equation}
H^\Lambda_N=h_0 + e_\Lambda h_1 +e_\Lambda^2 h_2 +
\cdot\cdot\cdot + e_\Lambda^N h_N \;.
\end{equation}
Having obtained the hamiltonian to some order in $e_\Lambda$, the next step
is to split it into two parts,
\begin{equation}
H^\Lambda=\H_0+\V \;.
\end{equation}
$\h0$ must be accurately solved
non-perturbatively, producing a zeroth order approximation for the
eigenvalues and eigenstates.  The greatest ambiguities in the
calculation appear in the choice of $\h0$, which requires one of
science's most powerful computational tools, trial and error.

In QED and QCD I {\it conjecture} that for sufficiently small $\Lambda$ all
interactions in
$\H_0$ preserve particle number, with all interactions that involve particle
creation and annihilation in $\V$. Corrections from $\V$ are then computed
in bound state perturbation theory.

Since $\H_0$ is assumed to include interactions that preserve particle
number, the zeroth order positronium ground state is a pure
electron-positron state.  We only need  one- and two-body interactions; {\it
i.e.}, the electron self-energy and the electron-positron interaction.  The
hamiltonian is computed to second order using Eq. (\ref{ham2}).  We must
specify $s_0$ in Eq. (\ref{ham2}), which corresponds to the inverse cutoff
squared at which boundary conditions are placed on the hamiltonian. Coupling
coherence leads to the prescription that $s_0 \rightarrow 0$ for irrelevant
operators and $s_0 \rightarrow \infty$ for relevant operators.

Bare electron mixing with electron-photon states leads to a self-energy (see
Fig. 1):

\begin{eqnarray}
\Sigma^{\Lambda}_{coh}(p)&=& {e_\Lambda^2 \over 8\pi^2 p^+}
\Biggl\{2 y \Lambda^2 \ln\Biggl(
{ y^2 \Lambda^2 \over (y\Lambda^2 + m^2)\epsilon}\Biggr)
-{3 \over 2} y \Lambda^2+{1 \over 2}
{ym^2 \Lambda^2 \over y \Lambda^2+m^2} \nonumber
\\ &&~~~~~~~~~~~~~~~~~~~~~~~~+ 3 m^2
\ln\Biggl( {m^2 \over y \Lambda^2 + m^2} \Biggr) \Biggl\} + {\cal
O}(\epsilon/y) \;;
\end{eqnarray}
where $y$ is the fraction of longitudinal momentum carried by the electron,
$y=p^+/P^+$. To simplify the discussion I have replaced the gaussian cutoff
factors that appear in all integrals with step functions, and completed the
integrals analytically using $1/\sqrt{s} = \Lambda^2/P^+$. It is possible to
produce such step function cutoffs with a similarity
transformation,\cite{brazil} but this leads to pathologies at higher order.
More importantly, I have been forced to introduce {\it a second cutoff},
\begin{equation}
x p^+ > \epsilon P^+ \;,
\end{equation}
because there is a logarithmic divergence in the loop longitudinal momentum
integration even with the gaussian cutoff in place. This
second cutoff must be taken to zero and no new counterterms can be added to
the hamiltonian, so all divergences must cancel before it is taken to zero.

We have no
choice about whether this divergent operator is in the hamiltonian if we use
coupling coherence.  We can only choose between putting it in $\H_0$ or in
$\V$.  I make different choices in QED and QCD, and the arguments are based
on physics.

The {\it divergent} electron `mass' is a complete lie.  We encounter
a term proportional to $e_\Lambda^2 \Lambda^2 \ln(1/\epsilon)/P^+$
when the scale is $\Lambda$; however, we can reduce this scale as far
as we please in perturbation theory.  Photons are massless, so the
electron continues to dress itself with small-x photons to
arbitrarily small $\Lambda$. Since I believe that this divergent
self-energy is exactly canceled by mixing with small-x photons, and
that this mixing is perturbative in QED, I simply put
it in $\V$.

There are two time-ordered diagrams involving photon exchange
between an electron with initial momentum $p_1$ and final momentum
$p_2$, and a positron with initial momentum $k_1$ and final momentum
$k_2$.  These are shown in Fig. 5, along with the instantaneous
exchange diagram. I refer the reader to longer articles where details are
given \cite{brazil,BJ97a} and concentrate here on the essential
results.

\begin{figure}
\epsfxsize=20pc
\center{\epsfbox{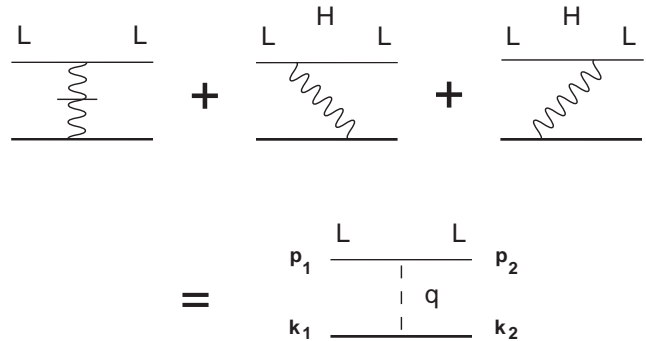}}
\caption{Effective two-body interaction between low energy
constituents resulting from: (i) the canonical instantaneous exchange
interaction, and (ii) the elimination of direct coupling between
low energy two-body states and high energy states containing an
additional gauge particle.}
\label{fig:fig5}
\end{figure}

Photon exchange above the cutoff produces an effective interaction
that cancels the instantaneous photon exchange operator in the canonical
hamiltonian, replacing it with a Coulomb interaction and magnetic
interactions that are partially responsible for fine structure. Instantaneous
photon exchange below the cutoff remains, and as is discussed in the section
on QCD it produces a long-range confining interaction. In QED this long range
interaction is exactly cancelled by further exchange of massless low energy
photons in bound state perturbation theory. In QCD the long range
interaction also acts on gluons, blocking such a cancellation and producing
the essential difference between QED and QCD in this approach.

This means that we can concentrate on photon exchange above the cutoff to
leading order, which still leaves us with a complicated operator in
light-front coordinates. In order to present an analytic analysis I
make assumptions that are justified $a~ posteriori$.  First I assume
that the electron and positron momenta can be arbitrarily large, but that in
low-lying states their relative momenta satisfy
\begin{equation}
|{\bf p}_\perp-{\bf k}_\perp| \sim \alpha m \;,
\end{equation}
\begin{equation}
|p^+-k^+| \sim \alpha (p^++k^+) \;.
\end{equation}
This allows us to use power counting to evaluate the
perturbative strength of operators for small coupling.

Given these order of magnitude estimates for momenta, we can
drastically simplify all of the operators in the hamiltonian.  At this point
we can complete the zeroth order analysis of positronium using the state,
\begin{eqnarray}
|\Psi(P)\rangle &=& \sum_{\sigma \lambda} \int {dp^+ d^2p_\perp \over
16\pi^3 p^+} {dk^+ d^2k_\perp \over 16\pi^3 k^+} \sqrt{p^+ k^+}
16\pi^3 \delta^3(P-p-k) \nonumber \\
&&~~~~~~~~~~~~~~\phi(p,\sigma;k,\lambda) b^\dagger(p,\sigma)
d^\dagger(k,\lambda) |0\rangle \;,
\end{eqnarray}
where $\phi(p,\sigma;k,\lambda)$ is the wave function for
the relative motion of the electron and positron, with the
center-of-mass momentum being $P$.  We need to choose the longitudinal
momentum appearing in the cutoff, and I use the natural scale
$P^+$.

If we want to find a cutoff for which the ground state is dominated by
the electron-positron component of the wave function, we need the
cutoff to remove the important part of the electron-positron-photon
phase space. Since the exchanged photon energy is typically ${\cal O}(\alpha
m^2)$, we need:
\begin{equation}
\Lambda^2<\alpha m^2 \;.
\end{equation}
On the other hand, we cannot allow the cutoff to remove the
region of the electron-positron phase space from which the wave
function receives most of its strength.  This requires
\begin{equation}
\Lambda^2>\alpha^2 m^2 \;.
\end{equation}

For cutoffs that satisfy $\alpha m^2 > \Lambda^2 > \alpha^2 m^2$, the
bound state equation can be simplified to:
\begin{equation}
-E \phi({\bf k}_1) = { {\bf k}_1^2 \over m} \phi({\bf k}_1)
-\alpha \int {d^3 k_2 \over (2 \pi)^3} {1 \over \bigl({\bf k}_1^2 -
{\bf k}_2^2\bigr)} \phi({\bf k}_2) \;,
\end{equation}
where a simple change of variables has replaced longitudinal momentum
fractions with a z-component of momentum, making the system's
nonrelativistic dynamics manifest. The cutoffs drop out to leading order,
leaving us with the familiar nonrelativistic Schr{\"o}dinger equation for
positronium in momentum space.  The solution is
\begin{equation}
\phi({\bf k}) = {{\cal N} \over \bigl({\bf k}^2+m E\bigr)^2} \;,
\end{equation}
\begin{equation}
E = {1 \over 4} \alpha^2 m \;.
\end{equation}
${\cal N}$ is a normalization constant.
This is the Bohr energy for the ground state of positronium, and it
is obvious that the entire nonrelativistic spectrum is reproduced to
leading order.

Beyond this leading order result the calculations become much more
interesting, and in any hamiltonian formulation they rapidly become
complicated.  The leading correction to the binding energy is
$\order(\alpha^4)$, and producing these corrections is a much more serious
test of the renormalization procedure. We have shown
that the fine structure of positronium is correctly reproduced when the
first- and second-order corrections from bound state perturbation theory are
added.\cite{BJ97a}  This is a formidable calculation, because the exact
Coulomb bound and scattering states appear in second-order bound state
perturbation theory.

A complete calculation of the Lamb shift in hydrogen would also
require a fourth-order similarity calculation of the hamiltonian;
however, the dominant contribution to the Lamb shift that was first
computed by Bethe \cite{bethe} can be computed using a hamiltonian
determined to $\order({\alpha})$.\cite{BJ97c}  In our calculation a
Bloch transformation was used rather than a similarity transformation
because the Bloch transformation is simpler and small energy
denominator problems can be avoided in analytical QED calculations.

The primary obstacle to using our light-front strategy for precision
QED calculations is algebraic complexity.  We have successfully used
QED as a testing ground for this strategy, but these calculations can
be done much more conveniently using other methods.  The theory for
which we believe our methods are best suited is QCD.

\subsection{Light-front QCD}

We only require the QCD hamiltonian determined to $\order(\alpha)$ to
discuss a simple confinement mechanism that appears naturally in
light-front QCD and to complete reasonable zeroth order calculations
for heavy quark bound states.  To this order the QCD hamiltonian in
the quark-antiquark sector is almost identical to the QED hamiltonian
in the electron-positron sector.  Of course the QCD hamiltonian
differs significantly from the QED hamiltonian in other sectors, and
this is essential for justifying my choice of $\H_0$ for
non-perturbative calculations.

The basic strategy for doing a sequence of (hopefully) increasingly
accurate QCD bound state calculations is almost identical to the
strategy for doing QED calculations.  Find an expansion for $H^\Lambda$ in
powers of the QCD coupling constant to a finite order.  Divide the
hamiltonian into a non-perturbative part, $\h0$, and a perturbative part,
$\V$.  The division is based on the physical argument that adding a parton
in an intermediate state should require more energy than indicated by the
free hamiltonian, and that as a result these states `freeze out'
as the cutoff approaches $\Lambda_{QCD}$. When this happens the
evolution of the hamiltonian as the cutoff is lowered further changes
qualitatively, and operators that were consistently canceled over an
infinite number of scales also freeze, so that their effects in the
few parton sectors can be studied directly.  A one-body operator and
a two-body operator arise in this fashion, and serve to confine both
quarks and gluons.

The simple confinement mechanism I outline is certainly not the final
story, but it may be the seed for the full confinement mechanism.  One
of the most serious problems we face when looking for non-perturbative
effects such as confinement is that the search itself depends on the
effect.  A candidate mechanism must be found and then shown to
produce itself self-consistently as the cutoff is lowered towards
$\Lambda_{QCD}$.

Once we find a candidate confinement mechanism, it is possible
to study heavy quark bound states with little modification of the QED
strategy.  Of course the results in QCD differ from those in
QED because of the new choice of $\H_0$, and in higher orders
because of the gluon interactions.

When we compute the QCD hamiltonian to $\order(\alpha)$, several
significant new features appear.  First are the familiar gluon
interactions.  In addition to the many gluon interactions found in the
canonical hamiltonian, there are modifications to the instantaneous
gluon exchange interactions, just as there were modifications to the
electron-positron interaction.  For example, a Coulomb interaction
automatically arises at short distances.  In addition the gluon self-energy
differs drastically from the photon self-energy.

The photon develops a self-energy because it mixes with
electron-positron pairs, and this self energy is $\order(\alpha
\Lambda^2/P^+)$.  When the cutoff is lowered below $4 m^2$, this mass
term dies exponentially because it is no longer possible to produce
electron-positron pairs.  For all cutoffs the small bare
photon self-energy is exactly canceled by mixing
with pairs below the cutoff.  I do not go through the
calculation, but because the gluon also mixes with gluon pairs in
QCD, the gluon self-energy acquires an infrared divergence, just as
the electron did in QED.  In QCD both the quark and gluon
self-energies are proportional to $\alpha \Lambda^2
\ln(1/\epsilon)/P^+$, where $\epsilon$ is the secondary cutoff on
parton longitudinal momenta introduced in the last section.  This
means that even when the primary cutoff $\Lambda^2$ is finite, the
energy of a single quark or a single gluon is infinite, because we are
supposed to let $\epsilon \rightarrow 0$.

In QED I argued that the bare electron self-energy is a complete lie,
because the bare electron mixes with photons carrying arbitrarily small
longitudinal momenta to cancel this bare self-energy and produce a
finite mass physical electron.  However, in QCD there is no reason to
believe that this perturbative mixing continues to arbitrarily small
cutoffs.  There are {\it no} massless gluons in the world. In this
case, the free QCD hamiltonian is a complete lie and cannot be
trusted at low energies.

On the other hand, coupling coherence gives us no choice about the
quark and gluon self-energies as computed in perturbation theory.
The question is not whether large
self-energies appear in the hamiltonian.  The question is whether
these self-energies are canceled by mixing with low energy
multi-gluon states.  As the
cutoff approaches $\Lambda_{QCD}$, I speculate that these
cancellations cease to occur because perturbation theory breaks
down and a mass gap between states with and without extra gluons
appears.

But if the quark and gluon self-energies diverge, and the divergences
cannot be canceled by mixing between sectors with an increasingly
large number of partons, how is it possible to obtain finite mass
hadrons?  The parton-parton interaction also diverges, and the
infrared divergence in the two-body interaction {\it exactly cancels} the
infrared divergence in the one-body operator for color singlet
states.

Of course, the cancellation of infrared divergences is not enough to
obtain confinement.  The cancellation is exact regardless of the
relative motion of the partons in a color singlet state, and
confinement requires a residual interaction.  The
$\order(\alpha)$ QCD hamiltonian contains a logarithmic potential in
both longitudinal and transverse directions.  There is no rigorous
demonstration that the confining interaction is linear, and a logarithmic
potential is of interest phenomenologically for heavy quark
bound states.\cite{quigg} I would be delighted if a
better light-front calculation produces a linear potential, but this
may not be necessary even for successful light hadron calculations.

The calculation of how the quark self-energy changes when a similarity
transformation lowers the cutoff on energy transfer is almost
identical to the electron self-energy calculation. We find the one-body
operator required by coupling coherence,
\begin{eqnarray}
\Sigma^{\Lambda}_{coh}(p)&=&
{g^2 C_F \over 8\pi^2 p^+} \Biggl\{2 y \Lambda^2 \ln\Biggl({ y^2
\Lambda^2 \over (y \Lambda^2+m^2) \epsilon }
\Biggr) -{3 \over 2} y
\Lambda^2+{1 \over 2} {y m^2 \Lambda^2 \over y \Lambda^2+m^2} \nonumber \\
&&~~~~~~~~~~~~~~~+ 3 m^2
\ln\Biggl( {m^2 \over y \Lambda^2 + m^2} \Biggr) \Biggl\} + {\cal
O}(\epsilon/y) \;,
\end{eqnarray}
where $C_F=(N^2-1)/(2N)$ for a SU(N) gauge theory.

The calculation of the quark-antiquark interaction required
by coupling coherence is also nearly identical to the QED
calculation.
Just as in QED the coupling coherent interaction induced by
gluon exchange above the cutoff partially cancels instantaneous gluon
exchange.  For the discussion of confinement the part of $V_{coh}$
that remains is not important, because it produces the short
range part of the Coulomb interaction.  However, the part of the
instantaneous interaction that is not canceled is
\begin{eqnarray}
\tilde{V}^\Lambda_{instant} &=& - 8 g_{\Lambda}^2 C_F
\sqrt{p_1^+ p_2^+ k_1^+ k_2^+}
\Biggl({1 \over q^+}\Biggr)^2 \delta_{\sigma_1 \sigma_2}
\delta_{\lambda_1 \lambda_2} \nonumber \\
&&\times
\;\;\theta\bigl(|p_1^+-p_2^+|-\epsilon P^+\bigr) \nonumber \\
&&\times \;\;{\rm exp}\Biggl[-2 s (p_1^--p_2^--q^-) (k_2^--k_1^--q^-) \Biggr]
\;.
\end{eqnarray}

Note that this interaction contains a cutoff that projects onto
exchange energies below the cutoff, because the interaction has been
screened by gluon exchange above the cutoffs. This interaction can
become important at long distances, if parton exchange below the
cutoff is dynamically suppressed.  In QED I argued that this singular
long range interaction is exactly canceled by photon exchange below
the cutoff, because such exchange is not suppressed no matter how low
the cutoff becomes.  Photons are massless and experience no
significant interactions, so they are exchanged to arbitrarily low
energies as effectively free photons.  This cannot be the case for
gluons.

For the discussion of confinement, place the most singular
parts of the quark self-energy and the quark-antiquark interaction in
$\h0$.  To see that all infrared divergences cancel and that the
residual long range interaction is logarithmic, study the
matrix element of these operators for a quark-antiquark state,
\begin{eqnarray}
|\Psi(P)\rangle &=& \sum_{\sigma \lambda} \sum_{rs}
\int {dp^+ d^2p_\perp \over
16\pi^3 p^+} {dk^+ d^2k_\perp \over 16\pi^3 k^+} \sqrt{p^+ k^+}
16\pi^3 \delta^3(P-p-k) \nonumber \\
&&~~~~~~~~~~~~~~\phi(p,\sigma,r;k,\lambda,s) b^{r\dagger}(p,\sigma)
d^{s\dagger}(k,\lambda) |0\rangle \;,
\end{eqnarray}
where $r$ and $s$ are color indices and
$\phi$ is a color singlet.

When the expectation value of the hamiltonian is taken using this state
there are divergences as $\epsilon \rightarrow 0$ in both the expectation
value of the self-energy and the surviving piece of instantaneous gluon
exchange. These divergences cancel exactly for any color-singlet state. The
cancellation resembles what happens in the Schwinger model. If the state is
a color octet the divergences are both positive and cannot cancel.  Since
the cancellation occurs in the matrix element, {we can let $\epsilon
\rightarrow 0$ before diagonalizing} $\H_0$.

The fact that the divergences cancel exactly does not indicate that
confinement occurs.  This requires the residual interactions to
diverge at large distances. It is easily shown that for large longitudinal
separations the interaction becomes:
\begin{eqnarray}
V(x^-)&=& {g_\Lambda^2 C_F \Lambda^2 \over 2\pi^2}  
\ln\bigl(|x^-|) \;.
\end{eqnarray}
At large transverse separations it becomes:
\begin{eqnarray}
V(x_\perp) &=& {g_\Lambda^2 C_F \Lambda^2 \over \pi^2} \ln(|{\bf x_\perp}|)
\;.
\end{eqnarray}
The strength of the long-range logarithmic
potential is not spherically symmetrical in these coordinates, with
the potential being larger in the transverse than in the longitudinal
direction.  Of course, there is no reason to demand that the
potential be rotationally symmetric in these coordinates, because rotations
are dynamical and are supposed to alter the Fock space composition of states
in addition to rotating the state in a given Fock space sector.

Had we computed the quark-gluon or gluon-gluon interaction, we would
find essentially the same residual long range two-body interaction in
every Fock space sector, although the strengths would differ because
different color operators appear.  In QCD gluons have a
divergent self-energy and experience divergent long range
interactions with other partons if we use coupling coherence.  In
this sense, the assumption that gluon exchange below some cutoff is
suppressed is consistent with the hamiltonian that results from this
assumption.  To show that gluon exchange is suppressed when $\Lambda
\rightarrow \Lambda_{QCD}$, rather than some other scale ({\it i.e.},
zero as in QED), a non-perturbative calculation of gluon exchange is
required. This same confinement mechanism appears using the similarity
transformation developed by Walhout.\cite{walhout}

I provide only a brief summary of our heavy quark bound
state calculations,\cite{Br96a,Br97a} and refer the reader to the
original articles for details.  We follow the strategy that has been
successfully applied to QED, with modifications suggested by the fact
that gluons experience a confining interaction.

For heavy quark bound states \cite{Br97a} we can simplify the hamiltonian by
making a nonrelativistic reduction and solving a Schr{\"o}dinger equation. 
We must then choose values for $\Lambda$, $\alpha$, and $M$.  These should be
chosen differently for bottomonium and charmonium. The cutoff for which the
constituent approximation works well depends on the constituent mass, as in
QED where it is obviously different for positronium and muonium.  In order
to fit the ground state and first two excited states of charmonium, we use
$\Lambda=2.5 GeV$, $\alpha=0.53$, $M_c=1.6 GeV$.  In order to fit these
states in bottomonium we use $\Lambda=4.9 GeV$, $\alpha=0.4$, and $M_b=4.8
GeV$.  Violations of rotational invariance from the remaining parts of the
potential are only about 10\%, and we expect corrections from higher Fock
state components to be at least of this magnitude for the couplings we use.

These calculations show that the approach is reasonable, but they are
not yet very convincing.  There are a host of additional calculations
that must be done before the success of this approach can be judged.

%%%%%%%%%%%%%%%%%%%%%%%%%%%%%%%%%%%%%%%%%%%%%%%%%%%%%%%%%%%%%%%%%%%%%%%%%%

%
\section*{Acknowledgments}
I wish to thank the organizers of the APCTP-RCNP Joint International
School on Physics of Hadrons and QCD for putting together an excellent
meeting. I would also like to acknowledge many useful discussions with Brent
Allen, Stan G{\l}azek, Roger Kylin, Rick Mohr, Jim Steele, and Ken Wilson.
This program has been advanced by many theorists, including Edsel Ammons,
Martina Brisudova, Koji Harada, Avaroth Harindranath, Billy Jones, Yizhang
Mo, Dave Robertson, Tim Walhout, and Wei-Min Zhang. This work was supported
by National Science Foundation grant PHY-9800964.

\end{document}